\documentstyle[sprocl,subfigure]{article}

\input{epsf}

\newcommand{\eqnref}[1]{Eq.~(\ref{#1})}
\newcommand{\eqnlessref}[1]{(\ref{#1})}

\def\Re{{\rm Re}}
\def\Im{{\rm Im}}
\def\Det{{\rm Det}}
\def\scrL{{\cal L}}
\def\scrD{{\cal D}}

\def\negspace{\kern -0.4em}
\def\ehat{{\bf \hat e}}

\begin{document}

%\preprint{UW/PT-00-11}

\title{DUAL QCD, EFFECTIVE STRING THEORY, AND REGGE TRAJECTORIES}

\author{M. Baker, R. Steinke}

\address{%
	Department of Physics,
	University of Washington, \\
	Seattle, WA 98195-1560 \\
	E-mail: baker@phys.washington.edu
	}%

\maketitle\abstracts{%
We start with an effective field theory containing classical
vortex solutions and show that the fluctuations of these vortices are
described by an effective string theory.  Viewed as a model for long
distance QCD, this theory provides a concrete picture of the QCD string as
a fluctuating Abrikosov-Nielsen-Olesen vortex of a dual superconductor on
the border between type I and type II.  We present arguments which suggest
that the action of the effective string theory is the Nambu-Goto action,
i.e. the rigidity vanishes.  We then use this theory to calculate the
corrections to classical Regge trajectories due to string fluctuations.
%We start with an effective field theory, dual QCD, containing classical
%vortex solutions, and obtain an effective string theory of
%these vortices. We present arguments which suggest that
%the action of the effective string theory is the
%Nambu--Goto action, and use this theory to calculate
%corrections to classical Regge trajectories due to
%string fluctuations.
}

\section{Introduction}

This talk is dedicated to the memory of
Fredrik Zachariasen, a close friend and
colleague of one of us (MB) for over forty years.
The work we will describe today was inspired by a
fifteen year collaboration with Fred
and Jim Ball. Fred invented the
name ``dual QCD,'' and the sound of
these words evokes a vivid
memory of Fred, and of how much
the collaboration meant to me.

We first review the results of dual QCD,
which is an effective field theory for long distance QCD
containing classical vortex
solutions. We then show that the fluctuations of
these superconducting vortices are described by
an effective string theory. We present
arguments which suggest that the action of this effective
string theory is the Nambu--Goto action. Finally, we
apply the results of a semi-classical expansion of the effective
theory to calculate corrections
to classical Regge trajectories due to string fluctuations.

\section{The Transformation from Fields to Strings}

\begin{figure}[ht]
	\epsfysize = 1.8in
	\leavevmode{\hfill \epsfbox{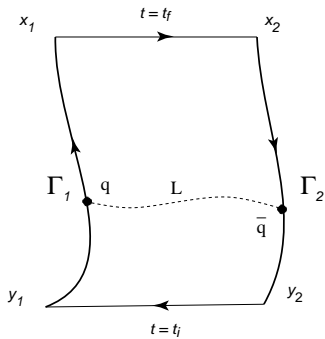} \hfill}
	\medskip
	\caption{The Loop $\Gamma$ \label{fig:Wilson loop}}
\end{figure}

In the dual superconductor picture of
confinement~\cite{Nambu}~\cite{Mand+tHooft}, a dual Meissner
effect confines electric color flux (${\bf Z_3}$~flux) to narrow
tubes connecting quark--antiquark pairs. Calculations
with a concrete version~\cite{Baker2}
of this model (dual QCD) has been compared both with
experimental data and with Monte Carlo simulations of
QCD~\cite{Bali}.
To a good approximation, the dual Abelian Higgs
model (with a suitable color factor) can be used to describe
these calculations. The Lagrangian $\scrL_{{\rm eff}}$
describing long distance QCD in the dual superconductor
picture then has the form:
\begin{equation}
\scrL_{{\rm eff}} = \frac{4}{3} \left\{ -\frac{1}{4} G_{\mu\nu}^2
- \frac{1}{2}\left|(\partial_\mu - igC_\mu)\phi\right|^2
- \frac{\lambda}{4}(|\phi|^2 - \phi_0^2)^2 \right\} \,.
\label{Leff}
\end{equation}
The potentials $C_\mu$ are dual potentials, and
$\phi$ is a complex Higgs field carrying monopole
charge, whose vacuum expectation value $\phi_0$ is
nonvanishing.
All particles are massive:
$M_\phi = \sqrt{2\lambda}\phi_0$, $M_C = g\phi_0$. The dual
coupling constant is $g = \frac{2\pi}{e}$, where $e$ is the
Yang--Mills coupling constant.
The potentials $C_\mu$ couple to the $q\bar q$ pair via $G_{\mu\nu}^S$,
a Dirac string whose ends are a source and a sink of electric color
flux,
\begin{equation}
G_{\mu\nu}^S = - e \int d^2\xi \frac{1}{2} \epsilon^{ab}
\epsilon_{\mu\nu\alpha\beta} \frac{\partial\tilde x^\alpha}{\partial\xi^a}
\frac{\partial\tilde x^\beta}{\partial\xi^b}
\delta^{(4)}\left( x^\mu - \tilde x^\mu(\xi) \right) \,.
\label{G string def}
\end{equation}
The field strength $G_{\mu\nu}$ is expressed in terms
of $C_\mu$ and $G_{\mu\nu}^S$,
\begin{equation}
G_{\mu\nu} = \partial_\mu C_\nu - \partial_\nu C_\mu + G_{\mu\nu}^S \,.
\label{G mu nu def}
\end{equation}
The effect of the string is to create a flux tube
(Abrikosov--Nielsen--Olesen (ANO) vortex~\cite{Abrikosov}) along some
line $L$ connecting the quark--antiquark pair, on which the
dual Higgs field $\phi$ must vanish. As the pair moves, the
line $L$ sweeps out a space time surface $\tilde x^\mu$,
whose boundary is the loop $\Gamma$ formed by the world lines
of the quark and antiquark trajectories. (See Fig. \ref{fig:Wilson loop})
The monopole field $\phi$
vanishes on the surface $\tilde x^\mu(\xi)$ parameterized
by $\xi^a$, $a=1,2$:
\begin{equation}
\phi(\tilde x^\mu(\xi)) = 0 \,.
\label{phi=0bc}
\end{equation}
\eqnref{phi=0bc} determines the location $\tilde x^\mu$ of the ANO
vortex of the field configuration $\phi(x^\mu)$.
The long distance $q\bar q$ interaction is determined by
the Wilson loop $W[\Gamma]$,
\begin{equation}
W[\Gamma] = \int \scrD C_\mu
\scrD\phi \scrD\phi^* e^{iS[C_\mu,\phi]} \,.
\label{Weff}
\label{originalpartition}
\end{equation}
The functional integration goes over
all field configurations containing a vortex sheet whose
boundary is $\Gamma$, and the action is
\begin{equation}
S[C_\mu,\phi] = \int d^4x \scrL_{{\rm eff}} \,.
\label{Seff}
\end{equation}

Previous calculations of $W[\Gamma]$ were  carried out in the classical
approximation (corresponding to a flat vortex sheet $\tilde x^\mu(\sigma)$),
and showed~\cite{Baker2} that the Landau--Ginzburg parameter $\lambda/g^2$
is approximately equal to $\frac{1}{2}$.
Both the vector and scalar particles have the same
mass $M=g\phi_0 \approx 910$ MeV,
and the flux tube radius is $a=\frac{\sqrt{2}}{M}$.

The classical approximation neglects the effect
of fluctuations in the shape of the flux tube
on the $q\bar q$ interaction. To take into account these
fluctuations, we must evaluate the functional
integral \eqnlessref{Weff} beyond the
classical approximation.
We carry out this integration in two steps:
(1) We fix the location of a vortex sheet $\tilde x^\mu$,
and integrate only over field configurations for which $\phi(x^\mu)$
vanishes on $\tilde x^\mu$. (2) We integrate
over all possible vortex sheets.
To implement this procedure, we introduce into the
functional integral \eqnlessref{originalpartition}
the factor one, written in the form
\begin{equation}
1 = J[\phi] \int \scrD \tilde x^\mu
\delta\left[\Re\phi(\tilde x^{\mu}(\xi))\right]
\delta\left[\Im\phi(\tilde x^{\mu}(\xi))\right] \,.
\label{inserttildex}
\end{equation}
The integration $\scrD \tilde x^\mu$ is over the four
functions $\tilde x^\mu(\xi)$. The functions $\tilde x^\mu(\xi)$
are a particular parameterization of the worldsheet $\tilde x^\mu$.

The expression \eqnlessref{inserttildex} implies that the string worldsheet
$\tilde x^\mu$, determined by the $\delta$ functions, is the
surface of the zeros of the field $\phi$. Inserting \eqnlessref{inserttildex}
into \eqnlessref{originalpartition} puts the Wilson loop in the form
\begin{equation}
\negspace
W[\Gamma] = \int \scrD\phi^* \scrD\phi \scrD C^{\mu} e^{iS[\phi,C]}
J[\phi] \int \scrD \tilde x^\mu
\delta\left[\Re\phi(\tilde x^{\mu}(\xi))\right]
\delta\left[\Im\phi(\tilde x^{\mu}(\xi))\right] \,.
\negspace
\label{beforeswitch}
\end{equation}
We then reverse the order of the field integration and the string
integration over surfaces $\tilde x^\mu(\xi)$,
\begin{equation}
\negspace
W[\Gamma] = \int \scrD \tilde x^\mu \int \scrD\phi^* \scrD\phi
\scrD C^{\mu} J[\phi] \delta\left[\Re\phi(\tilde x^{\mu}(\xi))\right]
\delta\left[\Im\phi(\tilde x^{\mu}(\xi))\right] e^{iS[\phi,C]} \,.
\negspace
\label{afterswitch}
\end{equation}
In \eqnref{beforeswitch}, the $\delta$ functions fix $\tilde x^\mu$
to lie on the surface of the zeros of a given field $\phi$, while in
\eqnref{afterswitch}, they restrict the field $\phi$ to vanish on a
given surface $\tilde x^\mu$. The integral over $\phi$ in
\eqnref{afterswitch} is therefore restricted to functions $\phi$ which
vanish on $\tilde x^\mu$, in contrast to the integral over $\phi$ in
\eqnref{beforeswitch}, in which $\phi$ can be any function.

\section{Factorization of the Jacobian}

\label{Jacobian factor}

The Jacobian $J[\phi]$ in \eqnref{afterswitch} is evaluated
for field configurations $\phi$ which vanish on a
particular surface $\tilde x^\mu$.
We make this explicit by writing \eqnlessref{inserttildex} as
\begin{equation}
J[\phi,\tilde x^{\mu}]^{-1} = \int \scrD \tilde y^\mu
\delta\left[\Re\phi(\tilde y^{\mu}(\tau))\right]
\delta\left[\Im\phi(\tilde y^{\mu}(\tau))\right] \,,
\label{Jdef}
\end{equation}
where $\tilde y^\mu$ is some other string worldsheet, distinct from
$\tilde x^\mu$.

The $\delta$ functions in \eqnlessref{Jdef} select
surfaces $\tilde y^\mu(\tau)$ which lie in a neighborhood of
the surface $\tilde x^\mu(\xi)$ of the zeros of $\phi$.
We separate
$\tilde y^\mu(\tau)$ into components lying on the
surface $\tilde x^\mu(\xi)$ and components lying along
vectors $n_\mu^A(\xi)$ normal to $\tilde x^\mu(\xi)$ at the point $\xi$:
\begin{equation}
\tilde y^\mu(\tau) = \tilde x^\mu(\xi(\tau)) + y_\perp^A(\xi(\tau))
n_\mu^A(\xi(\tau)) \,.
\label{localcoords}
\end{equation}
\begin{figure}[ht]
	\epsfxsize = 2in
	\leavevmode{\hfill \epsfbox{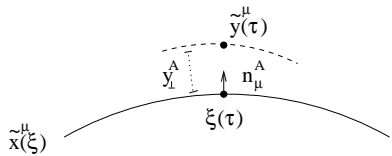} \hfill}
	\medskip
	\caption{Worldsheets and normal vectors \label{normal figure}}
\end{figure}
The point $\tilde x^\mu(\xi(\tau))$ is the point on the
surface $\tilde x^\mu(\xi)$ lying closest to $\tilde y^\mu(\tau)$,
and the magnitude of $y_\perp^A(\xi(\tau))$ is the distance
from $\tilde y^\mu(\tau)$ to $\tilde x^\mu(\xi(\tau))$
(see Fig.~\ref{normal figure}).

We now exhibit the factorization of the Jacobian.
Making the change of coordinates $\tilde y^\mu(\tau)
\to \left(\xi(\tau), y_\perp^A(\xi)\right)$ gives
\begin{equation}
J[\phi,\tilde x^{\mu}]^{-1} = \int \scrD \xi \scrD y_{\perp}^A
\Det_\tau[\sqrt{-g}] \delta\left[\Re\phi\left(\tilde x^{\mu} +
y_{\perp}^A n_A^{\mu}\right)\right]
\delta\left[\Im\phi\left(\tilde x^{\mu}
+ y_{\perp}^A n_A^{\mu}\right)\right] \,,
\label{J partial split}
\end{equation}
where $\sqrt{-g}$ is the square root of the determinant
of the induced metric
\begin{equation}
g_{ab} = \frac{\partial\tilde x^\mu}{\partial\xi^a}
\frac{\partial\tilde x^\mu}{\partial\xi^b} \,.
\end{equation}
\eqnref{J partial split} has the form:
\begin{equation}
J[\phi,\tilde x]^{-1} = \int \scrD \xi(\tau)
\Det_{\tau}\left[\sqrt{-g}\right] J_{\perp}[\phi,
\tilde x^{\mu}(\xi(\tau))]^{-1} \,,
\label{newJacob}
\end{equation}
where
\begin{eqnarray}
J_{\perp}[\phi,\tilde x^{\mu}]^{-1} &=& \int \scrD y_{\perp}^A
\delta\left[\Re\phi\left(\tilde x^{\mu} + y_{\perp}^A
n_A^{\mu}\right)\right]
\delta\left[\Im\phi\left(\tilde x^{\mu}
 + y_{\perp}^A n_A^{\mu}\right)\right]
\end{eqnarray}
contains all the dependence on $\phi$. The Jacobian $J_\perp$ is
the Faddeev-Popov determinant for the degrees of freedom
$y_\perp^A$ which move the string.

Since $J_\perp$ is
independent of the parameterization $\xi(\tau)$, the
Jacobian factors into two parts:
\begin{equation}
J[\phi,\tilde x]^{-1} = J_{\parallel}[\tilde x]^{-1} J_{\perp}[\phi,
\tilde x]^{-1} \,,
\label{Jfactor}
\end{equation}
where
\begin{equation}
J_{\parallel}[\tilde x]^{-1} = \int \scrD \xi
\Det_{\tau}\left[\sqrt{-g}\right] \,.
\label{Jparallel}
\end{equation}
The string part $J_\parallel$
of the Jacobian arises from the parameterization
degrees of freedom.
In the following section, we will use $J_\parallel$ to fix
the reparameterization degrees of freedom.

\section{Effective String Theory of Vortices}

\label{effective action section}

Inserting the factorized form \eqnlessref{Jfactor} of $J[\phi]$
into the expression \eqnlessref{afterswitch} for $W[\Gamma]$
gives the Wilson Loop the form
\begin{equation}
W[\Gamma] = \int \scrD \tilde x^\mu J_\parallel[\tilde x]
e^{iS_{{\rm eff}}} \,,
\label{stringrep}
\label{vortex partition}
\end{equation}
where the action $S_{{\rm eff}}$ of the effective string theory is
given by
\begin{equation}
e^{iS_{{\rm eff}}[\tilde x^\mu(\xi)]} = \int \scrD\phi^* \scrD\phi
\scrD C^\mu J_\perp[\phi] \delta\left[\Re\phi(\tilde x^\mu(\xi))\right]
\delta\left[\Im\phi(\tilde x^\mu(\xi))\right] e^{iS} \,.
\label{Sstring}
\end{equation}
The string action \eqnlessref{Sstring} was obtained previously by Gervais
and Sakita~\cite{Gervais+Sakita}. The novel feature of our
result is the string integration measure of the Wilson loop
\eqnlessref{stringrep}.

Any surface $\tilde x^\mu$ has only two physical degrees of
freedom. The other two degrees of freedom represent the invariance of
the surface under coordinate reparameterizations.
We fix the coordinate reparameterization symmetry by choosing
a particular ``representation'' $x^\mu$ of the surface, which
depends on two functions $f^1(\xi)$, $f^2(\xi)$,
\begin{equation}
x^\mu(\xi) = x^\mu[f^1(\xi), f^2(\xi), \xi] \,.
\label{x_p def}
\end{equation}

Any physical surface can be expressed in terms of $x^\mu$
by a suitable choice of $f^1$ and $f^2$. In particular, the
worldsheet $\tilde x^\mu(\xi)$
appearing in the integral \eqnlessref{vortex partition}
can be written in terms of a reparameterization $\tilde\xi(\xi)$
of the representation $x^\mu$,
\begin{equation}
\tilde x^\mu(\xi) = x^\mu[f^1(\tilde\xi(\xi)),
f^2(\tilde\xi(\xi)), \tilde\xi(\xi)] \,.
\label{x tilde of x_p}
\end{equation}
The four degrees of freedom in $\tilde x^\mu(\xi)$ are
replaced by two physical degrees of freedom $f^1(\xi)$, $f^2(\xi)$
and two reparameterization degrees of freedom $\tilde\xi(\xi)$.

We can write the integral over $\tilde x^\mu(\xi)$
in \eqnlessref{vortex partition} in terms of integrals over
$f^1(\xi),f^2(\xi)$ and $\tilde\xi(\xi)$,
\begin{equation}
\scrD \tilde x^\mu
= \Det\left[ \tilde t_{\mu\nu} \sqrt{-g}
\frac{\partial x^\mu}{\partial f^1} \frac{\partial x^\nu}
{\partial f^2} \right] \scrD f^1 \scrD f^2 \scrD \tilde\xi \,,
\label{measure shift}
\end{equation}
where
\begin{equation}
\tilde t_{\mu\nu} = \frac{1}{2} \epsilon_{\mu\nu\alpha\beta}
\frac{\epsilon^{ab}}{\sqrt{-g}} \frac{\partial x^\alpha}{\partial\xi^a}
\frac{\partial x^\beta}{\partial\xi^b} \,,
\end{equation}
is the antisymmetric tensor normal to the worldsheet.

With the parameterization \eqnlessref{x tilde of x_p}
of $\tilde x^\mu$,
the path integral \eqnlessref{vortex partition} takes the form
\begin{equation}
W[\Gamma] = \int \scrD\tilde\xi \scrD f^1 \scrD f^2
\Det\left[ \tilde t^{\mu\nu} \frac{\partial x^\mu}
{\partial f^1} \frac{\partial x^\nu} {\partial f^2} \right]
\Det[\sqrt{-g}] J_\parallel e^{iS_{{\rm eff}}} \,.
\label{W subbed}
\end{equation}
Due to the invariance of the theory under coordinate
reparameterizations, the only term in \eqnlessref{W subbed}
which depends on $\tilde\xi$ is the determinant
of $\sqrt{-g}$. When we bring the terms which are
independent of $\tilde\xi$ outside of the integral,
the path integral becomes
\begin{equation}
W[\Gamma] = \int \scrD f^1 \scrD f^2
\Det\left[ \tilde t^{\mu\nu} \frac{\partial x^\mu}
{\partial f^1} \frac{\partial x^\nu} {\partial f^2} \right]
J_\parallel e^{iS_{{\rm eff}}} \int \scrD \tilde\xi
\Det[\sqrt{-g}] \,.
\label{before cancel J parallel}
\end{equation}

The remaining integral over reparameterizations $\tilde\xi$
is equal to $J_\parallel^{-1}$, defined by \eqnlessref{Jparallel},
and is canceled by the explicit factor of $J_\parallel$ appearing
in \eqnlessref{before cancel J parallel}. This means we
do not need to evaluate $J_\parallel$, and can
avoid the complications inherent in evaluating the
integral over reparameterizations of the string
coordinates. The anomalies~\cite{Polyakov} produced in string theory
by evaluating this integral are not present, and
there is no Polchinski--Strominger term~\cite{Pol+Strom,ACPZ}
in the theory.  \eqnref{before cancel J parallel}
gives the final result for the Wilson loop
\begin{equation}
W[\Gamma] = \int \scrD f^1 \scrD f^2 \Det\left[ \tilde t_{\mu\nu}
\frac{\partial x^\mu}{\partial f^1} \frac{\partial x^\nu}
{\partial f^2} \right] e^{iS_{{\rm eff}}} \,,
\label{param measure}
\end{equation}
as an integration over two function $f^1(\xi)$ and $f^2(\xi)$,
the physical degrees of freedom of the string. ``Gauge fixing''
the reparameterization symmetry has produced a Faddeev-Popov
determinant.

\section{The Effective Action}

The action \eqnlessref{Sstring} of the effective string theory
gives the action $S_{{\rm eff}}(\tilde x^\mu)$
of the effective string theory \eqnlessref{param measure}
as an integral over field configurations
which have a vortex fixed at $\tilde x^\mu$. Since the vortex
theory \eqnlessref{originalpartition} is an effective long distance
theory, the path integral \eqnlessref{originalpartition} for $W[\Gamma]$,
written in terms of the fields of the Abelian Higgs model,
is cut off at a scale $\Lambda$ which is on the order of
the mass $M$ of the dual gluon. Furthermore, the
integration \eqnlessref{stringrep} over $\tilde x^\mu$
includes all the long distance fluctuations of the theory.
Therefore, the path integral \eqnlessref{Seff}
contains neither short distance nor long distance fluctuations,
and is determined by minimizing the field action
$S[\tilde x^\mu, \phi, C_\mu]$ for a fixed position of the vortex sheet:
\begin{equation}
S_{{\rm eff}}[\tilde x^\mu] = S[\tilde x^\mu]
\equiv S[\tilde x^\mu, \phi^{{\rm class}},
C_\mu^{{\rm class}}] \,, \kern 0.5 in \phi^{{\rm class}}(\tilde x^\mu) = 0 \,.
\label{classical string}
\end{equation}
The fields $\phi^{{\rm class}}$ and $C_\mu^{{\rm class}}$
are the solutions of the classical equations of motion,
subject to the boundary condition $\phi(\tilde x^\mu) = 0$.

\begin{figure}[ht]
	\epsfxsize = 2in
	\leavevmode{\hfill \epsfbox{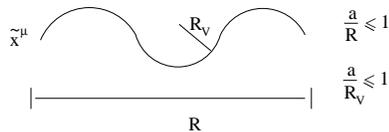} \hfill}
	\medskip
	\caption{Length scales of the string \label{fig:curvature}}
\end{figure}

To evaluate $W[\Gamma]$, we need to know the classical
action $S[\tilde x^\mu]$
for strings of length $R$ and radius of curvature $R_V$
greater than the flux tube radius $a$ (see Fig.~\ref{fig:curvature}).
In the case of long straight strings, $a/R \ll 1$, $a/R_V \ll 1$,
the classical action $S$ becomes the Nambu--Goto action $S_{{\rm NG}}$,
\begin{equation}
S[\tilde x^\mu] = S_{{\rm NG}} \equiv -\sigma \int d^2\xi \sqrt{-g} \,.
\label{NG action first}
\end{equation}

In the next two sections, we consider separately long bent
strings ($R\to\infty$) and short straight strings
($R_V\to\infty$). We present arguments that suggest that
the action \eqnlessref{NG action first}
is a good approximation to $S_{{\rm eff}}[\tilde x^\mu]$
in both these situations.

\section{Long Bent Strings; the Rigidity}

For long bent strings ($R\to\infty$), the leading correction
to the Nambu--Goto action is the curvature term
\begin{equation}
S_{{\rm curvature}} = - \beta \int d^2\xi \sqrt{-g}
\left(-\nabla^2 \tilde x^\mu\right)^2 \sim \frac{a^2}{R_V^2} S_{NG} \,,
\label{curvature term}
\end{equation}
where
\begin{equation}
-\nabla^2 \tilde x^\mu = - \frac{1}{\sqrt{-g}}
\frac{\partial}{\partial\xi^a} \sqrt{-g} g^{ab}
\frac{\partial}{\partial\xi^b} \tilde x^\mu \,.
\end{equation}
The calculation of the ``rigidity'' $\beta$ determining
the size of $S_{{\rm curvature}}$ has been considered by
a number of authors~\cite{Maeda}, but its value
for a superconductor on the I--II border was never calculated.
We conjecture that $\beta=0$ on the basis of analytic results
for an infinite Nielsen--Olesen flux tube on the
(I/II) border obtained by de Vega and Schaposnik~\cite{deVega+Schaposnik}.
We now briefly describe their results.

\begin{figure}[ht]
	\centering
	\mbox{
		\subfigure[\label{fig:straight}]{\epsfysize = 1.6in
			\epsfbox{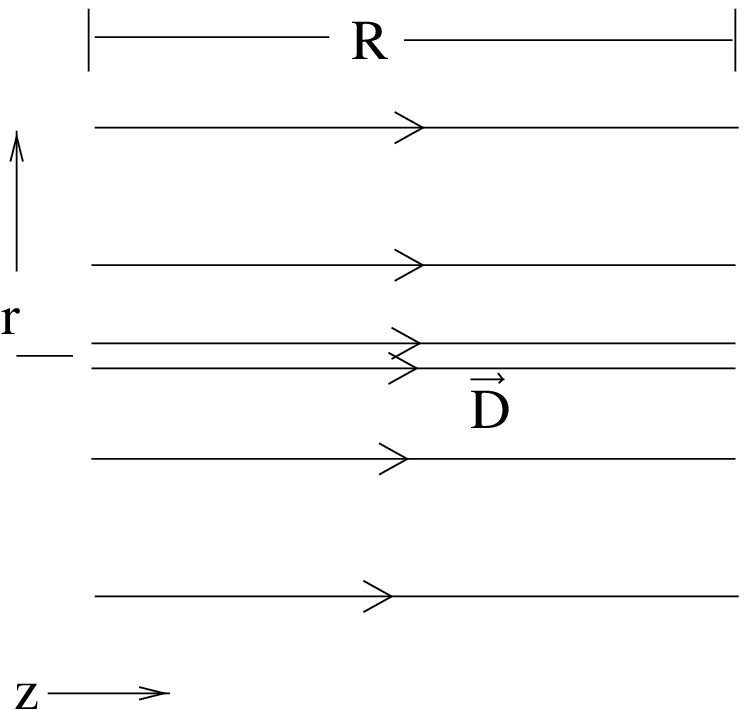}}\quad
		\subfigure[\label{fig:bent}]{\epsfysize = 1.6in
			\epsfbox{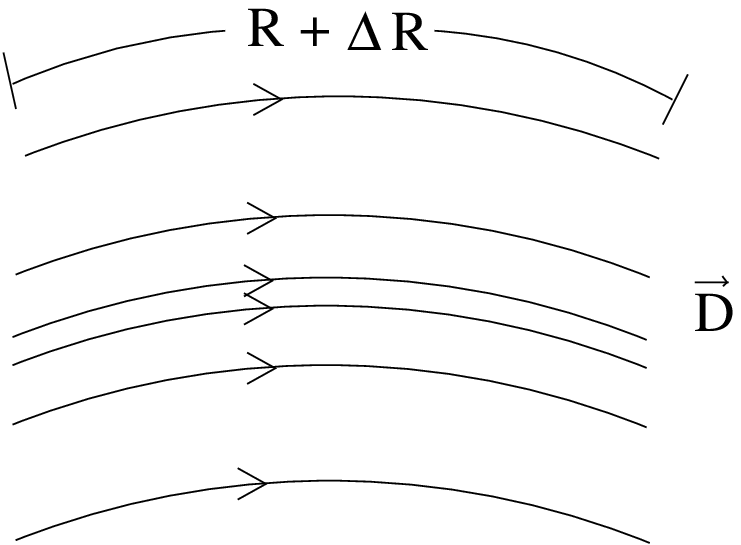}}\quad
	}
	\epsfxsize = 3.5in
%	\leavevmode{\hfill \epsfbox{combined.eps} \hfill}
	\medskip
	\caption{Field lines for (a) a straight
	and (b) a bent vortex \label{combined figure}}
\end{figure}

We denote the color field by $\vec D$:
\begin{equation}
\vec D = D_{{\rm FT}}(r) \ehat_z \,,
\kern 1 in
D_{{\rm FT}}(r) = G_{\theta r}(r) \,,
\label{vec D def}
\end{equation}
where $r$ and $\theta$ are the radial coordinates in a plane
perpendicular to the axis of the flux tube, which lies
along the $z$ axis. The
field $\vec D$ in a tube segment of length $R$ is indicated in
Figure~\ref{combined figure}.
We separate the action $S$ into a Higgs contribution $S_\phi$
and a gauge contributions $S_g$,
\begin{equation}
S = S_\phi + S_g \,,
\label{classical action split}
\end{equation}
where
\begin{equation}
S_\phi = \int d^4x \left[ - \frac{1}{2} \left( \partial_\mu |\phi|\right)^2
- \frac{\lambda}{4} \left( |\phi|^2 - \phi_0^2 \right)^2 \right] \,,
\label{Higgs part}
\end{equation}
and
\begin{equation}
S_g = \int d^4x \left[ - \frac{1}{2} G_{\mu\nu}^2
- \frac{1}{2} g^2 C_\mu^2 |\phi|^2 \right] \,.
\label{gauge part}
\end{equation}
For a straight flux tube,
\begin{equation}
S_\phi = -\sigma_\phi \int d^2\xi \sqrt{-g} \,,
\kern 1 in
S_g = -\sigma_g \int d^2\xi \sqrt{-g} \,,
\label{straight broken}
\end{equation}
and the string tension $\sigma$ is the sum $\sigma =
\sigma_\phi + \sigma_g$. Furthermore, the difference
\begin{equation}
S_g - S_\phi = \int d^4x T_{\theta\theta} \,,
\end{equation}
where $T_{\theta\theta}$ is the $\theta\theta$ component of
the stress tensor.

de Vega and Schaposnik showed that for a superconductor on
the (I/II) border
\begin{equation}
\sigma_\phi = \sigma_g = \sigma/2
= \frac{\pi}{2} \phi_0^2 \,,
\hbox{ so that }
S_\phi - S_g = 0 \,.
\end{equation}
They also showed the components of the stress tensor
in the plane perpendicular to the flux tube vanish,
$T_{rr} = T_{\theta\theta} = 0$.
The repulsion between the lines of force of the gauge field
is compensated by the attraction produced by the Higgs fields.
In other words, there are no ``bonds'' perpendicular to the axis
of a straight flux tube.

Now suppose the flux tube is bent slightly, so that $R \to R + \Delta R$
as indicated in Figure~\ref{combined figure}.
Since no perpendicular bonds are stretched or compressed, the
corresponding change in the energy should be the
string tension multiplied by the change in length.
That is, the curvature term, which in a sense represents the
attraction or repulsion between neighboring parts of the
string, should vanish.

We now present some formal arguments which support this
conjecture. We consider a general vortex sheet $\tilde x^\mu$
(see Fig.~\ref{fig:curvature}), and again separate the action
into a Higgs contribution \eqnlessref{Higgs part}
and a gauge contribution \eqnlessref{gauge part}.
Using the classical equations of motion for the gauge
field $C^\mu$, we can write the action $S_g$
in the alternate form
\begin{equation}
S_g = -\frac{1}{4} \int d^4x G_{\mu\nu}^S G^{\mu\nu} \,.
\label{gauge action near singular}
\end{equation}
Since $G_{\mu\nu}^S$ is nonzero only on the worldsheet
$\tilde x^\mu$, \eqnref{gauge action near singular}
can be written as an integral over the worldsheet
$\tilde x^\mu(\xi)$. The field tensor $G_{\mu\nu}$ is a
function of the spacetime point $x^\mu$ and a
functional of the worldsheet $\tilde x^\mu$.
We define the color field $D(\xi,\tilde x^\mu)$
at the point $x^\mu = \tilde x^\mu(\xi)$ on the
worldsheet by the equation
\begin{equation}
\frac{1}{2} D^2(\xi,\tilde x^\mu) = \frac{1}{4} G_{\mu\nu}^2
(x^\mu, \tilde x^\mu) \Big|_{x^\mu = \tilde x^\mu(\xi)} \,.
\end{equation}
\eqnref{gauge action near singular} for $S_g$ can then be
written in the form
\begin{equation}
S_g = -\frac{e}{2} \int d^2\xi \sqrt{-g} D(\xi,\tilde x^\mu) \,.
\label{S_g of D}
\end{equation}

We now find the change in the action when we vary
the position of the worldsheet $\tilde x^\mu$. Since
the classical solution is a stationary point of the
action, only the explicit dependence of $S$ on $\tilde x^\mu$
contributes to $\delta S$. This explicit dependence
is only present in the term $\frac{1}{4} \int d^4x G_{\mu\nu}^2$
which contains $G_{\mu\nu}^S$. Hence, when $\tilde x^\mu \to
\tilde x^\mu + \delta\tilde x^\mu$, then $S \to S + \delta S$,
where
\begin{equation}
\delta S = S[\tilde x^\mu + \delta\tilde x^\mu] - S[\tilde x^\mu]
= -\frac{1}{2} \int d^4x \delta G_{\mu\nu}^S G^{\mu\nu} \,.
\end{equation}
Alternately, we can use \eqnlessref{classical action split} and
\eqnlessref{gauge action near singular} to evaluate $\delta S$
explicitly,
\begin{equation}
\delta S = \delta S_\phi - \frac{1}{4} \int d^4x \delta G_{\mu\nu}^S
G^{\mu\nu} - \frac{1}{4} \int d^4x G_{\mu\nu}^S \delta G^{\mu\nu} \,,
\end{equation}
or
\begin{equation}
-\frac{1}{4} \int d^4x G_{\mu\nu}^S \delta G^{\mu\nu}
= \frac{1}{2} \delta S - \delta S_\phi
= \frac{1}{2} \delta \left( S_g - S_\phi \right) \,.
\label{combined result}
\end{equation}
Writing \eqnlessref{combined result} in terms of $\delta D$ gives
\begin{equation}
e \int d^2\xi \sqrt{-g} \frac{\delta D(\xi, \tilde x^\mu)}
{\delta\tilde x^\mu} = -\frac{\delta\left( S_g - S_\phi \right)}
{\delta\tilde x^\mu} \,.
\label{delta D found}
\end{equation}
\eqnref{delta D found} relates the the change in the
color field $D$ on the string to the change of the difference
$S_g - S_\phi$ as the position $\tilde x^\mu$ of the worldsheet
is varied.

In the limit of a flat sheet, $D(\xi,\tilde x^\mu)$
is a constant independent of the position $\xi$ on the
worldsheet,
\begin{equation}
D(\xi,\tilde x^\mu) = D_{{\rm FT}} \equiv D_{{\rm FT}}(r=0) \,,
\label{D_FT first}
\end{equation}
and \eqnref{S_g of D} for $S_g$ reduces to the Nambu--Goto
action with $\sigma_g = \frac{e}{2} D_{FT}$.
We can estimate the size of the corrections to
\eqnlessref{D_FT first} for slightly bent sheets, by
using \eqnlessref{straight broken} to evaluate the right hand side
of \eqnlessref{delta D found}. This gives
\begin{eqnarray}
e \int d^2\xi \sqrt{-g} \frac{\delta D(\xi,\tilde x^\mu)}
{\delta\tilde x^\mu} &\approx& (\sigma_g - \sigma_\phi)
\frac{\delta}{\delta\tilde x^\mu} \int d^2\xi \sqrt{-g}
\nonumber \\
&=& (\sigma_g - \sigma_\phi) \int d^2\xi \sqrt{-g}
\left(-\nabla^2 \tilde x^\mu\right) \,.
\label{vary get curvature}
\end{eqnarray}
The quantity $-\nabla^2 \tilde x^\mu$ is proportional to
the extrinsic curvature, which is of order $a/R_V$,
so that the correction to the flat sheet limit
\eqnlessref{D_FT first} is of order $a/R_V$.
However, for a superconductor on the (I--II) border,
$\sigma_g - \sigma_\phi = 0$, so that the flat sheet
expressions \eqnlessref{straight broken} and \eqnlessref{D_FT first}
for $S_\phi$, $S_g$, and $D(\xi,\tilde x^\mu)$ are
compatible with the relations \eqnlessref{S_g of D} and
\eqnlessref{delta D found} for sheets with nonvanishing
extrinsic curvature. This suggests that the
rigidity vanishes. \eqnref{vary get curvature} is the formal
manifestation of the heuristic argument given earlier that bending
a flux tube on the (I/II) border does not produce a perturbation
of the order $a/R_V$.

\section{Short Straight Strings}

Recent numerical studies~\cite{Gubarev+Polikarpov+Zakharov}
of the classical equations for a flat sheet have
shown that for a superconductor on the border
the Nambu--Goto action \eqnlessref{NG action first}
remains a good approximation for short straight
strings ($a/R \le 1$, $R_V = \infty$).
We now show that this result is compatible with
equations \eqnlessref{S_g of D} and \eqnlessref{delta D found},
written in a form appropriate to a straight string
of finite length $R$. In this situation, the
action $S = TV(R)$, where $T$ is the elapsed time and
$V(R)$ is the static potential between a $q \bar q$ pair
separated by a distance $R$. We divide $V(R)$ into a Coulomb
part $V^{{\rm Coulomb}}$ and a nonperturbative part
$V^{{\rm NP}}$:
\begin{equation}
V(R) = V^{{\rm Coulomb}}(R) + V^{{\rm NP}}(R) \,.
\label{V of R break}
\end{equation}
We then separate the gauge and Higgs contributions to $V^{{\rm NP}}(R)$,
\begin{equation}
V^{{\rm NP}}(R) = V_\phi + V_g^{{\rm NP}} \,.
\end{equation}
Then \eqnref{S_g of D} becomes:
\begin{equation}
V^{{\rm NP}}_g = \frac{e}{2} \int_{-R/2}^{R/2} dz D^{{\rm NP}}(z,R) \,,
\label{V_g of D}
\end{equation}
where $D^{{\rm NP}}(z,R)$ is the nonperturbative part
of the color field at a point $z$ on the string.
The change $\delta \tilde x^\mu$ in the string
is replaced by the change $\delta R$ in its length,
so \eqnref{delta D found} becomes
\begin{equation}
e \int_{-R/2}^{R/2} dz \frac{\partial D^{{\rm NP}}(z,R)}{\partial R}
= \frac{d}{dR} \left[ V^{{\rm NP}}_g(R) - V_\phi(R) \right] \,.
\label{partial of D}
\end{equation}
At large $R$, the nonperturbative fields become the fields of an
infinitely long flux tube, so that
\begin{equation}
D^{{\rm NP}}(z,R) \to D_{{\rm FT}}\,,  \hbox{ for }
-\frac{R}{2} \le z \le \frac{R}{2} \,,
\kern 0.4 in
V^{{\rm NP}}_g \to \sigma_g R \,,
\kern 0.4 in
V_\phi \to \sigma_\phi R \,.
\label{large R potentials}
\end{equation}
Using the large $R$ potentials on the right hand side of
\eqnlessref{partial of D} gives
\begin{equation}
\frac{e}{2} \int_{-R/2}^{R/2} dz \frac{\partial D^{{\rm NP}}(z,R)}
{\partial R} \approx \left( \sigma_g - \sigma_\phi \right) \,.
\label{limit partial D}
\end{equation}
\eqnref{limit partial D} shows that the color field $D^{{\rm NP}}(z,R)$
at a given point $z$ on the string depends upon the length $R$ of the
string. Therefore, the field on the interior of the
string can, in general, no longer remain a constant, $D_{{\rm FT}}$,
as $R$ becomes smaller.
However, for a superconductor on the (I/II) border
the right hand side of \eqnref{limit partial D} vanishes,
so that the solution $D^{{\rm NP}}(z,R) = D_{{\rm FT}}$
is also consistent at smaller $R$. In fact, the potential $V^{{\rm NP}}_g$,
which is the product of the field on the string and the length
$R$ of the string, can remain proportional to $R$ for
smaller values of $R$. The same is true for
$V_\phi(R)$, and the potential $V^{{\rm NP}} = \sigma R$
is consistent with the constraint \eqnlessref{partial of D}
for all values of $R$.

\section{Regge Trajectories for Light Mesons}

In this section we calculate corrections to classical Regge
trajectories due to string fluctuations.
We take the action of the effective string theory to be the
Nambu--Goto action for all $R$ and $R_V$ such that
$a/R \le 1$ and $a/R_V \le 1$,
\begin{equation}
S_{{\rm eff}} = -\sigma \int d^2\xi \sqrt{-g} \,.
\label{NG action}
\end{equation}
Combining \eqnlessref{param measure} and \eqnlessref{NG action}
gives the Wilson loop $W[\Gamma]$ of the effective string
theory,
\begin{equation}
W[\Gamma] = \int \scrD f^1 \scrD f^2 \Det\left[ \tilde t_{\mu\nu}
\frac{\partial x^\mu}{\partial f^1} \frac{\partial x^\nu}{\partial f^2}
\right] e^{-i\sigma \int d^2\xi \sqrt{-g}} \,.
\end{equation}
Let $\Gamma$ be
the loop generated by the worldlines of an equal mass
quark--antiquark pair separated by a distance $R$,
and rotating with angular velocity $\omega$. The
velocity of the quarks is $v = \omega R/2$.
The effective Lagrangian $L$ for the quark--antiquark
pair is given by
\begin{equation}
L(R,\omega) = -2m\sqrt{1-v^2} + L^{{\rm string}}(R,\omega) \,,
\label{L cl def}
\end{equation}
where
\begin{equation}
L^{{\rm string}} = \frac{-i}{T} \log W[\Gamma] \,.
\end{equation}
Since both $R$ and
$\omega$ are fixed, $L^{{\rm string}}$ is time independent.

We evaluate $L^{{\rm string}}$ by carrying out a semiclassical
expansion of $W[\Gamma]$ about the classical rigid rotating string
solution $\bar x^\mu$. This expansion gives
\begin{equation}
L^{{\rm string}} = L^{{\rm string}}_{{\rm cl}}
+ L_{{\rm fluc}} \,,
\label{L string divide}
\end{equation}
where
\begin{equation}
L^{{\rm string}}_{{\rm cl}} = -\frac{\sigma}{T} \int d^2\xi
\sqrt{-\bar g} \,,
\end{equation}
and $L_{{\rm fluc}}$ is the contribution of the long wavelength
transverse vibrations of the rotating string, which is
analogous to the result~\cite{Luscher}
of L\"uscher for static strings.

We can write \eqnlessref{L cl def} and \eqnlessref{L string divide}
in the form
\begin{equation}
L(R,\omega) = L_{{\rm class}} + L_{{\rm fluc}} \,,
\label{L eff}
\end{equation}
with
\begin{equation}
L_{{\rm class}} = -2m\sqrt{1-v^2} - \sigma \int_{-R/2}^{R/2}
dr \sqrt{1-r^2\omega^2} \,.
\end{equation}
The effective Lagrangian \eqnlessref{L eff} determines the
angular momentum $J$ and the energy $E$ of the rotating
quarks,
\begin{equation}
J = \frac{\partial L}{\partial\omega} \,,
\kern 1in
E = \omega \frac{\partial L}{\partial\omega} - L \,.
\end{equation}
The ``equation of motion'' $\frac{\partial L}{\partial R} = 0$
determines the frequency of rotation $\omega$ as a function of $R$,
$\omega = \omega(R)$.
We have calculated $L_{{\rm fluc}}$, and find that
$L_{{\rm fluc}} \ll L_{{\rm cl}}^{{\rm string}}$ for
large $R$. The semiclassical expansion is then justified,
and $L_{{\rm fluc}}$ can be treated as a
perturbation. The energy $E(J)$ is then related to
the energy $E_{{\rm class}}(J)$, calculated in
the absence of string fluctuations, by the equation
\begin{equation}
E(J) = E_{{\rm class}}(J) - L_{{\rm fluc}} \,.
\label{E perturbed}
\end{equation}
\eqnref{E perturbed} gives the Regge trajectory
\begin{equation}
J = \frac{E^2}{2\pi\sigma} - \sqrt{\frac{E}{\pi^3 m}} \left[
\ln\left(\frac{Mm}{\sigma}\right) + 1 \right]
- \frac{4}{3\sigma} \sqrt{\frac{m^3 E}{\pi}} + \frac{7}{12}
+ O\left( \frac{\ln E}{\sqrt{E}} \right) \,.
\label{J result}
\end{equation}
The short wavelength fluctuations are cut off at
a wavelength $\lambda \sim a \sim 1/M$.

\begin {figure}[ht]
    \begin{center}
	\begin{tabular}{rc}
	    \vbox{\hbox{$J$ \hskip 0in \null} \vskip 1in} &
	    \epsfxsize = 2.5in
	    \epsfbox{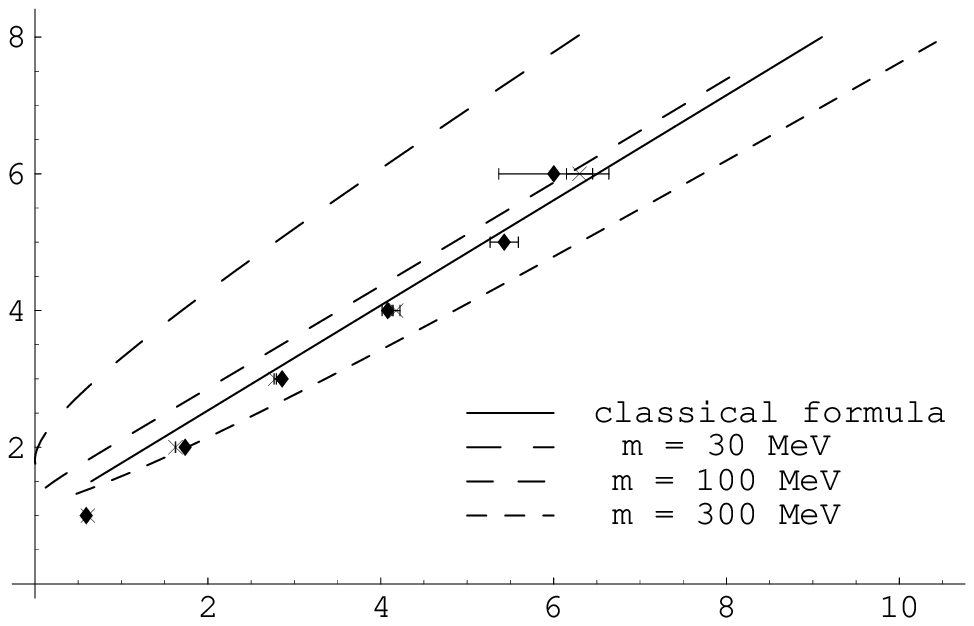} \\
	    &
	    \hbox{$E^2$ in GeV$^2$} \\
	\end{tabular}
    \end{center}
    \medskip
    \caption{$\rho$--$\omega$ Regge trajectory}
\label{regge figure}
\end {figure}

In Fig.~\ref{regge figure}, we plot the Regge trajectory
\eqnlessref{J result} using the values $M = 910$ MeV and
$\sigma = \left(455 \hbox{ MeV}\right)^2$ obtained previous
fits of heavy quark potentials~\cite{Baker2}.
We have chosen quark masses of 30, 100, and 300 MeV. For
comparison, we also plot the classical formula
$J = E^2/2\pi\sigma$. The plotted points are observed
particles on the $\omega$ and $\rho$ trajectories.
We have added one to the value of $J$ on the plot
to account for the spin of the quarks. We have chosen the
range of values for the quark masses in Fig.~\ref{regge figure}
in order to give a qualitative picture of the dependence of
the Regge trajectory on the quark mass. Since \eqnlessref{J result}
does not include the contribution of quark fluctuations to the
Regge trajectory, this formula is incomplete.

\newpage

\section{Conclusions}

\begin{enumerate}
\item The dual superconductor description of long distance QCD
yields the effective string theory \eqnlessref{param measure}
of superconducting vortices.
\item We have presented arguments which suggest
that the action of the effective
string theory is the Nambu--Goto action \eqnlessref{NG action first}.
\item The semiclassical expansion of the effective string
theory, including the fluctuations of the vortex,
gives the result \eqnlessref{J result} for
Regge trajectories.
We are in the process of taking into account the
quantum fluctuations of the quark degrees
of freedom in order to complete
the calculation of the semiclassical corrections
to Regge trajectories.
\end{enumerate}

\section*{Acknowledgments}

We would like to thank N. Brambilla for many
very helpful discussions, and for her important
contributions to the work presented here.
This work was supported in part
by the U. S. Department of Energy grant
DE-FG03-96ER40956.

\end{document}